\documentclass[man,12pt,nolmodern,floatsintext,hidelinks]{apa7}
\usepackage[nodisplayskipstretch]{setspace}
\usepackage{pdfpages}
\usepackage{newtxtext,newtxmath} %TNR font
\usepackage{booktabs,multirow,array}
\usepackage{graphicx}
\usepackage{enumitem}
\usepackage[nosectionbib,numberedbib]{apacite} 
\usepackage{url}
\usepackage{amsmath,amsfonts}
\usepackage{algorithm,algpseudocode}
\usepackage{tikz}
\usepackage{afterpage,lscape}
\usetikzlibrary{shapes,arrows.meta,decorations.markings,positioning}
\usepackage{bm} %bold math
\newcolumntype{L}[1]{>{\raggedright\let\newline\\\arraybackslash\hspace{0pt}}m{#1}}
\newcolumntype{C}[1]{>{\centering\let\newline\\\arraybackslash\hspace{0pt}}m{#1}}
\newcolumntype{R}[1]{>{\raggedleft\let\newline\\\arraybackslash\hspace{0pt}}m{#1}}

\DeclareMathAlphabet{\mathbf}{OT1}{cmr}{b}{n}

\definecolor{red1}{RGB}{228,55,55}
\definecolor{blue1}{RGB}{0,0,0}
\definecolor{coquelicot}{rgb}{1.0, 0.22, 0.0}
\long\def\jp#1{\bgroup\color{coquelicot}#1\egroup}
\long\def\yl#1{\bgroup\color{blue1}#1\egroup}

\def\t{^\prime}

\def\eeta{\boldsymbol{\eta}}

\def\xxi{\boldsymbol{\xi}}

\def\ee{\mathbf{e}}
\def\ss{\mathbf{s}}
\def\tt{\mathbf{t}}

\def\uu{\mathbf{u}}
\def\vv{\mathbf{v}}
\def\yy{\mathbf{y}}

\def\real{\mathbb{R}}

\def\var{\hbox{Var}}

\def\cov{\hbox{Cov}}
\def\corr{\hbox{Corr}}

\def\ex{\mathbb{E}}
\def\pr{\mathbb{P}}

\def\ind{\mathbb{I}}

\algrenewcommand\algorithmicrequire{\textbf{input:}}
\algrenewcommand\algorithmicensure{\textbf{output:}}
\algblockdefx{Do}{EndDo}{\textbf{do}}{\textbf{end do}}

\setlist[itemize, 1]{leftmargin=*, topsep=0ex, itemsep=0pt, parsep=0ex, labelindent=0pt}
\setlist[itemize, 2]{leftmargin=*, topsep=0ex, itemsep=0pt, parsep=0ex, labelindent=0pt, label=$\circ$}
\setlist[itemize, 3]{leftmargin=*, topsep=0ex, itemsep=0pt, parsep=0ex, labelindent=0pt, label=-}
\setlist[enumerate, 1]{leftmargin=*, topsep=0ex, itemsep=0pt, parsep=0ex, labelindent=0pt, label=(\alph*)}

\begin{document}
% title page
\title{On a General Theoretical Framework of Reliability}

\shorttitle{Reliability}
\authorsnames[1,2,3]{Yang Liu, Jolynn Pek, \& Alberto Maydeu-Olivares}
\authorsaffiliations{{Department of Human Development and Quantitative Methodology\\ University of Maryland, College Park},
{Department of Psychology, The Ohio State University},
{Department of Psychology, University of South Carolina and \\Faculty of Psychology, University of Barcelona}}
\authornote{Correspondence should be made to Yang Liu at
3304R Benjamin Bldg, 3942 Campus Dr, University of Maryland, College Park, MD
20742. Email: yliu87@umd.edu.}

% abstract and keywords
\abstract{%
\indent
  Reliability is an essential measure of how closely observed scores represent latent scores (reflecting constructs), assuming some latent variable measurement model. We present a general theoretical framework of reliability, placing emphasis on measuring the association between latent and observed scores. This framework was inspired by McDonald's \citeyear{mcdonald.2011} regression framework, which highlighted the coefficient of determination as a measure of reliability. We extend McDonald's \citeyear{mcdonald.2011} framework beyond coefficients of determination and introduce four desiderata for reliability measures (estimability, normalization, symmetry, and invariance). We also present theoretical examples to illustrate distinct measures of reliability and report on a numerical study that demonstrates the behavior of different reliability measures. We conclude with a discussion on the use of reliability coefficients and outline future avenues of research.
  }

\maketitle
\setcounter{secnumdepth}{0}

% introduction
%\centerline{\textbf{Introduction}}
Psychological theories are often developed and assessed using the notion of constructs. Constructs (e.g., attitudes, personality, psychopathy) cannot be directly observed and are often defined operationally as latent variables (LVs; \citeNP{hoyle.borsboom&tay.2024}; see also \citeNP{deboeck.et.al.2023a} for a recent discussion on the notion of constructs). LVs, and more generally functions of LVs, which we term \emph{latent scores}, are assumed to be reflected by manifest variables (MVs; e.g., item responses). \emph{Observed scores} that are functions of MVs are often computed to serve as proxies of latent scores to make inferences about constructs. In the developments to follow, we assume that constructs are validly operationalized in the population by an LV measurement model (e.g., item response theory [IRT] model [\citeNP{thissen&steinberg.2009}]), which formally expresses the link between MVs and LVs.\footnote{We recognize that our use of words ``variable'' and ``score'' are not fully aligned with their common usage in English. They should be treated as special terminologies throughout the paper. In particular, we refer to MVs by $\yy_i$ and LVs by $\eeta_i$ along with the subscript $i$ to denote each case. Observed and latent scores are respective functions of MVs and LVs.}

Observed scores (e.g., summed scores and estimated factor scores) are often employed for scoring, classification, and examining relations among constructs (e.g., see \citeNP{liu&pek.inpress}). When employing observed scores in research, it is pertinent to consider the extent to which observed scores map well onto latent scores that quantify psychological constructs. An imperfect mapping manifests as measurement error and might result in misleading inference (\citeNP{bollen.1989}, Chapter 5; \citeNP{cole&preacher.2014}). Thus, it is important to assess how well observed scores align with latent scores, which is gauged by reliability coefficients.

  Many popular reliability coefficients can be interpreted as coefficients of determination based on regression models (\citeNP{mcdonald.2011}; see \citeNP{liu.pek&maydeu-olivares.2024} for a review). For example, classical test theory (CTT) reliability is the coefficient of determination associated with regressing an observed score onto all LVs (in the measurement model), which is referred to as a \emph{measurement decomposition} of the observed score. CTT reliability quantifies how well these LVs account for variance of the observed score (e.g., \citeNP{anastasi&urbina.1997}; \citeNP{devellis&thorpe.2021}; \citeNP{raykov&marcoulides.2011}). Conversely, proportional reduction in mean squared error (PRMSE; \citeNP{haberman&sinharay.2010}) is the coefficient of determination associated with regressing a latent score onto all MVs (in the measurement model), which is referred to as a \emph{prediction decomposition} of the latent score. PRMSE is a popular measure of reliability in the IRT literature and indicates the proportion of latent score variance accounted for by MVs.

  The purpose of this paper is to extend the regression framework of reliability (\citeNP{mcdonald.2011}; see also \citeNP{liu.pek&maydeu-olivares.2024}), from which we derive novel reliability coefficients that also quantify the alignment between latent and observed scores. We frame reliability coefficients within the broader context of association measures, which include the coefficient of determination from the special case of the univariate regression framework \cite{liu.pek&maydeu-olivares.2024}. To organize new reliability coefficients under the extended framework, we introduce four desiderata, discuss several example reliability coefficients, and illustrate their behavior with a numerical study.

  The paper is organized as follows. We begin by introducing notation and preliminary concepts. Next, we briefly review the regression framework of reliability, focusing on the measurement and prediction decompositions that result in CTT reliability and PRMSE, respectively. We then consider reliability coefficients as measures of association between latent and observed scores, expanding the regression framework. To organize reliability coefficients under this generalized framework, we introduce four desiderata. The first two (estimability and normalization) are necessary whereas the next two (symmetry and invariance) are not essential. We then present five theoretical examples to illustrate the generality of the proposed framework: (a) squared Pearson's correlation \cite{kim.2012}, (b) coefficient sigma \cite{schweizer&wolff.1981}, (c) mutual information \cite{joe.1989, markon.2023}, (d) coefficient $T$ \cite{azadkia&chatterjee.2021}, and \yl{(e) a generalized coefficient of determination for multivariate regression  (i.e., coefficient $W$; cf. \citeNP{mardia&kent&bibby.1979}). The use of coefficients of (b), (d), and (e) in quantifying reliability is novel.} Next, we report on a numerical study investigating the performance of these reliability coefficients under a two-dimensional independent-cluster IRT model. Finally, we end with a discussion on limitations and future avenues of research.

  \centerline{\textbf{Reliability from a Regression Framework}}

  \noindent \textbf{Notation and Assumptions}\\

  Let $\yy_i$ be an $m \times 1$ vector of MVs for person $i$, in which $i=1, \dots, n$. The MVs are assumed to reflect LVs for person $i$ as represented by the $d \times 1$ vector $\eeta_i$. We also assume a correctly specified measurement model that formally links the MVs to the LVs, resulting in a joint probability density function (pdf) of $\underline\yy_i$ and $\underline\eeta_i$, denoted by $f(\yy_i, \eeta_i)$.\footnote{In the most general scenario, $\underline\yy_i$ and $\underline\eeta_i$ may combine continuous and discrete random variables. Therefore, the pdf should be understood as the Radon-Nikodym derivative with respect to a product measure that is composed of Lebesgue measures for continuous variates and counting measures for discrete variates.} \yl{Variables and vectors are underlined when they need to be highlighted as random.} Furthermore, let $\ss(\yy_i)$ denote a $m^*\times 1$ vector of observed scores and $\xxi(\eeta_i)$ denote a $d^*\times 1$ vector of latent scores. Here, $m^* \leq m$ and $d^*\leq d$. \yl{Examples of observed scores include summed scores \cite{sijtsma&ellis&borsboom.2024}, factor scores in common factor analysis \cite{bartlett.1937, thomson.1936, anderson&rubin.1956, mcdonald.1981}, and IRT scale scores \cite{thissen&wainer.2001}. In addition to LVs themselves, commonly used latent scores include CTT true scores and their percentile ranks \cite<e.g.,>{livingston&lewis.1995, lord.1980}.} While parameters to a measurement model are estimated from data in practice, we limit our discussion to focus on reliability measures in the population.

  \noindent \textbf{Reliability Coefficients Based on Regressions}\\

  Inspired by \citeA{mcdonald.2011}, \citeA{liu.pek&maydeu-olivares.2024} interpreted reliability coefficients as coefficients of determination based on univariate regressions. The measurement decomposition regresses a univariate observed score $s(\yy_i)$ (i.e., $m^* = 1$) onto all the LVs in $\eeta_i$. Conversely, the prediction decomposition regresses a univariate latent score $\xi(\eeta_i)$ (i.e., $d^* = 1$) onto all the MVs in $\yy_i$. As described below, the measurement decomposition yields CTT reliability and the prediction decomposition yields PRMSE.

  The measurement decomposition, defined for a scalar-valued observed score $s(\yy_i)$ and also known as the true score formula \cite<e.g.,>[Section 5.2]{raykov&marcoulides.2011}, can be expressed as
\begin{equation}
  s(\yy_i) = \ex\big[s(\underline \yy_i)|\eeta_i\big] + \varepsilon_i.
\label{eq:measdecomp}
\end{equation}
  Because a regression traces the conditional expectation of an outcome variable given explanatory variables (e.g., \citeNP{fox.2015}, p. 15), Equation \ref{eq:measdecomp} can be considered a (potentially nonlinear) regression of the observed score $s(\yy_i)$ onto the LVs in $\eeta_i$. The conditional expectation of $s(\underline \yy_i)$ given $\eeta_i$ is often referred to as the \emph{true score} underlying $s(\yy_i)$, and the error term $\underline\varepsilon_i$ has mean 0 and is uncorrelated with the true score $\ex\big[s(\underline \yy_i)|\eeta_i\big]$ \cite[Theorem 2.7.1]{lord&novick.1968}. Alternatively, Equation \ref{eq:measdecomp} can be viewed as a unit-weight linear regression (i.e., with intercept 0 and slope 1) of the observed score $s(\yy_i)$ onto its true score $\ex\big[s(\underline\yy_i)|\eeta_i\big]$. The corresponding coefficient of determination in Equation \ref{eq:measdecomp} quantifies the proportion of observed score variance that is explained by latent (true) score variance; this coefficient of determination is identical to CTT reliability:
\begin{equation}
    \varrho^2(s(\underline\yy_i), \underline\eeta_i) = \varrho^2\big(s(\underline\yy_i), \ex\big[s(\underline \yy_i)|\underline\eeta_i\big]\big) = \frac{\var\big(\ex\big[s(\underline\yy_i)|\underline\eeta_i\big]\big)}{\var\big[s(\underline\yy_i)\big]} = 1 - \frac{\ex\big(\var\big[s(\underline\yy_i)|\underline\eeta_i\big]\big)}{\var\big[s(\underline\yy_i)\big]}.
\label{eq:cttrel}
\end{equation}
In Equation \ref{eq:cttrel}, $\varrho^2(\underline u, \underline \vv)$ refers to the population coefficient of determination when regressing a scalar outcome variable $u$ onto (possibly multiple) explanatory variables $\vv$.
%For example, $\varrho^2(s(\underline\yy_i), \underline\eeta_i)$ refers to the population coefficient of determination from regressing $\eeta_i$ onto $\ss(\underline\yy_i)$.
The last equality is due to the \yl{law of total variance: $\var\big[s(\underline\yy_i)\big]$ = $\ex\big(\var\big[s(\underline\yy_i)|\underline\eeta_i]\big)$ + $\var\big(\ex\big[s(\underline\yy_i)|\underline\eeta_i\big]\big)$}. \yl{Coefficients omega and alpha are popular examples of CTT reliability \cite{cronbach.1951, mcdonald.1999}, and both coefficients are defined for summed scores. Coefficient omega assumes a congeneric measurement model (i.e., a common factor model with a single LV), whereas coefficient alpha assumes a more restrictive tau-equivalent model (i.e., a congeneric model with equal factor loadings).\footnote{\yl{It is also common to interpret coefficient alpha as a lower bound of CTT reliability, which holds under very weak assumptions \cite[Theorem 4.4.3]{lord&novick.1968}.}}}

  The prediction decomposition is defined for a scalar-valued latent score $\xi(\eeta_i)$ and is expressed as
\begin{equation}
    \xi(\eeta_i) = \mathbb{E}\big[\xi(\underline\eeta_i)|\yy_i\big] + \delta_i.
\label{eq:preddecomp}
\end{equation}
Equation \ref{eq:preddecomp} can also be interpreted in terms of two regressions. It is a (potentially nonlinear) regression of the latent score $\xi(\eeta_i)$ on all the MVs in $\yy_i$, or a unit-weight linear regression of $\xi(\eeta_i)$ on $\mathbb{E}\big[\xi(\underline\eeta_i)|\yy_i\big]$, which is the expected \textit{a posteriori} (EAP) score of $\xi(\eeta_i)$. Note that the EAP score minimizes the mean squared error (MSE) among all predictors of $\xi(\eeta_i)$, and the minimized MSE is given by $\ex\big(\var\big[\xi(\underline\eeta_i)|\underline\yy_i\big]\big)$.\footnote{\yl{Consider predicting $\xi(\eeta_i)$ by an observed score $s(\yy_i)$. The MSE in prediction is given by $\ex[ s(\underline\yy_i) - \xi(\underline\eeta_i)]^2$, which is minimized when $s(\yy_i) = \ex[\xi(\underline\eeta_i)|\yy_i]$, the EAP score of $\xi(\eeta_i)$. See \citeA[Exercise 4.13]{casella&berger.2002} for a justification.}} Thus, the coefficient of determination resulting from Equation \ref{eq:preddecomp} is
\begin{equation}
  \varrho^2(\xi(\underline\eeta_i), \underline\yy_i) = \varrho^2\big(\xi(\underline\eeta_i),\mathbb{E}\big[\xi(\underline\eeta_i)|\underline\yy_i\big]\big) = \frac{\var\big(\ex\big[\xi(\underline\eeta_i)|\underline\yy_i\big]\big)}{\var\big[\xi(\underline\eeta_i)\big]} = 1 - \frac{\ex\big(\var\big[\xi(\underline\eeta_i)|\underline\yy_i\big]\big)}{\var\big[\xi(\underline\eeta_i)\big]},
\label{eq:prmse}
\end{equation}
  which quantifies the proportion of MSE reduction when predicting $\xi(\eeta_i)$ from $\yy_i$. Equation \ref{eq:prmse}, henceforth termed PRMSE, is a popular measure of reliability in the IRT literature.

In sum, reliability coefficients are strictly defined as coefficients of determination within the regression framework \cite{liu.pek&maydeu-olivares.2024, mcdonald.2011}. In the measurement decomposition of an observed score, the explanatory variables must be all the LVs in the measurement model (or equivalently the true score underlying the observed score). Alternatively, in the prediction decomposition of a latent score, the explanatory variables must be all the MVs in $\yy_i$ (or equivalently the EAP predictor of the latent score). The coefficient of determination quantifies the magnitude of association between outcome and explanatory variables. Next, we extend the definition of reliability to more general measures of association between selected observed and latent scores.

\centerline{\textbf{Reliability as a Measure of Association}}

\yl{Let $A\big(\underline\uu, \underline\vv\big)$ be an association measure that maps a pair of random vectors $\underline\uu$ and $\underline\vv$ to a real number.\footnote{\yl{As a prerequisite, if the association measure $A\big(\underline\uu^*, \underline\vv^*\big)$ is defined for a specific pair of $\underline\uu^*\in\real^{a}$ and $\underline\vv^*\in\real^{b}$ for integers $a, b>0$, then the measure should be defined for all pairs of random vectors $\underline\uu\in\real^{a}$ and $\underline\vv\in\real^{b}$ of the same dimensions.}} The larger the value of the association measure, the more closely $\underline\uu$ and $\underline\vv$ are aligned in the population.  We define reliability by applying the association measure to the $m^*$-dimensional random observed score vector $\ss(\underline\yy_i)$ and the $d^*$-dimensional random latent score vector $\xxi(\underline\eeta_i)$; that is,
\begin{equation}
  A\big(\ss(\underline\yy_i), \xxi(\underline\eeta_i)\big).
\label{eq:genrel}
\end{equation}
Because CTT reliability and PRMSE are coefficients of determination, they are special cases of the general definition in Equation \ref{eq:genrel}. For CTT reliability, the observed score is $s(\yy_i)\in\real$, and the latent score(s) are either the LVs $\eeta_i\in\real^d$ or the true score $\ex\big[s(\underline\yy_i)|\eeta_i\big]\in\real$. For PRMSE, the observed score(s) are either the MVs $\yy_i\in\real^m$ or the EAP score $\ex\big[\xi(\underline\eeta_i)|\yy_i\big]\in\real$ and the latent score is $\xi(\eeta_i)\in\real$.}

\yl{Because our definition of reliability (Equation \ref{eq:genrel}) is completely general, we next discuss some desirable statistical properties that could be satisfied by the association measure. These desiderata facilitate the estimation and interpretation of reliability coefficients. Moreover, these desiderata serve as criteria to organize existing coefficients and might also be adopted as guiding principles to define new coefficients. The four desiderata are estimability, normalization, symmetry, and invariance. Estimability and normalization are unequivocally necessary properties. Estimability guarantees that we can accurately estimate reliability coefficients from sample data and appropriately quantify the sampling error (at least in large samples). Normalization ensures that the reliability coefficients are defined on a familiar and intuitive scale. In contrast to the two aforementioned desiderata, symmetry and invariance might only be desirable in certain contexts and thus are less essential.}

\yl{Our desiderata are motivated by but are less restrictive than the well-known ``R\'enyi's Axioms'' \cite<e.g., >{geenens&ldm.2022, nelsen.2006, renyi.1959, schweizer&wolff.1981}. R\'enyi's Axioms collect advisable statistical principles that define a specific class of association measures termed \emph{measures of dependence}. Our definition of reliability coefficients, however, are not confined to measures of dependence. We present a version of R\'enyi's Axioms in the Supplementary Materials.}

\noindent
\textbf{Estimability}\\

Recall that the joint distribution of $\underline\yy_i$ and $\underline\eeta_i$ is determined by the specified LV measurement model. Then, a reliability coefficient $A\big(\ss(\underline\yy_i), \xxi(\underline\eeta_i)\big)$ is a function of parameters in the measurement model and is thus also a population parameter. Estimability means that $A\big(\ss(\underline\yy_i), \xxi(\underline\eeta_i)\big)$ can be consistently estimated from an independent and identically distributed sample $\yy_1, \dots, \yy_n$, and that approximate confidence intervals (CIs) for $A\big(\ss(\underline\yy_i), \xxi(\underline\eeta_i)\big)$ can be constructed.

Consistent estimation of reliability coefficients can be assured under two conditions. First, the measurement model should be (locally) identified for model parameters to be consistently estimated \cite<see, e.g.,>[Chapter 2]{bekker.merckens&wansbeek.2014}. Second, $A\big(\ss(\underline\yy_i), \xxi(\underline\eeta_i)\big)$ should be an almost surely continuous function of the model parameters such that consistent estimates of reliability coefficients can be obtained by the continuous mapping theorem \cite<e.g.,>[Theorem 2.3]{vandervaart.1998}. Under complex nonlinear measurement models in which computations for reliability become intractable, we can approximate reliability coefficients using a large Monte Carlo (MC) sample of observed and latent scores generated from the fitted measurement model; see \yl{the Supplementary Materials} and \citeA{liu.pek&maydeu-olivares.2024} for details.

Large-sample CIs for reliability coefficients require additional assumptions. Analytical methods \cite<e.g., the Delta method;>[Section 5.3]{bickel&doksom.2015} are useful when efficient evaluation of model-implied quantities is viable. Resampling methods \cite<e.g., bootstrapping;>{efron&tibshirani.1993} are more convenient to implement due to their plug-and-play nature.

%In sum, estimability ensures consistent estimation of reliability coefficients, accompanied by large-sample CIs. %Due to space limit, we do not further consider interval estimation in the current review; however, its cruciality in small-sample studies should not be understated.

\noindent
\textbf{Normalization}\\
A normalized measure of association $A\big(\underline\ss(\yy_i), \xxi(\underline\eeta_i)\big)$ is defined on the unit interval $[0, 1]$. Normalization aids in interpretation because the value of zero indicates \textit{absence of association} and the value of one indicates \textit{perfect association}. In this vein, zero reliability implies that the observed scores contain only measurement error and are not relevant to the latent scores. Conversely, a value of one on reliability implies that observed scores are essentially equivalent to latent scores. Stated differently, the observed scores are free of measurement error and are perfect proxies of the latent scores.

The absence of association (zero reliability) has at least two interpretations. First, from the regression framework, a zero coefficient of determination implies that the conditional expectation of the outcome variable given the predictor variables has no variability. Stated differently, the conditional and unconditional expectations of the outcome are equal, sometimes referred to as linear independence \cite<e.g.,>[Definition 2.11.1]{lord&novick.1968}.
  %In a simple linear regression, a zero Pearson correlation suggests linear independence.
  Second, a zero coefficient can imply statistical independence; that is, the joint pdf of observed scores $\ss(\underline\yy_i)$ and latent scores $\xxi(\underline\eeta_i)$ can be factorized into the product of their marginal pdfs. Statistical independence implies linear independence but not vice versa. %For example, it is possible to find two random variables that are nonlinearly related (and thus statistically dependent) but have a zero Pearson correlation \cite<see, e.g., Example 4.5.9 of>{casella&berger.2002}.
  %commented out the example because it seems rather straightforward to readers of BJMSP.

  A perfect association implies a deterministic relationship between latent and observed variables. Different measures of association differ in (a) whether the deterministic relationship should be established in one direction or in both directions, and (b) which family of deterministic functions are involved. As for (a), the regression framework is asymmetric and a perfect association therein only requires the outcome to be a deterministic function of the explanatory variables. In contrast, a perfect symmetric association (see section below) implies that both sets of scores can be interchangeably represented as deterministic functions of each other. In terms of (b), families of deterministic functions include linear functions with nonzero slopes (e.g., the squared or absolute Pearson correlation), strictly monotone functions \cite<e.g., >{nelsen.2006, schweizer&wolff.1981}, and implicitly defined functions \cite<e.g.,>{geenens&ldm.2022}.

  While it is desirable to use scores with reliability close to one, it is challenging to suggest a universal cutoff of acceptable reliability for two reasons. First, different association measures are often not directly comparable, in which values on $[0, 1]$ might map onto qualitatively different concepts (e.g., we cannot compare measures with different conceptual definitions of zero and perfect associations). %To enhance interpretability, it has been recommended that association measures correspond to a strictly monotone function of the squared Pearson correlation when both the observed and latent scores are unidimensional and follow a bivariate normal distribution \cite<e.g.,>{geenens&ldm.2022, renyi.1959, schweizer&wolff.1981}. This additional requirement preserves the most familiar ``variance explained'' interpretation of reliability. \jp{[Commented this out because it didn't seem essential and it was a little rambly~]}
  Second, the same amount of measurement error may have different downstream effects depending on the use of observed scores \cite<e.g., recovering latent scores, classifying individuals, and being entered as proxies of latent scores in an explanatory model; see >{liu&pek.inpress}. There is no shortcut but to study the consequences of measurement error in a case-by-case fashion. We will revisit this point in the ``Numerical Study'' section with a concrete example.

\noindent
\textbf{Symmetry}\\

The association measure (Equation \ref{eq:genrel}) is \textit{symmetric} if and only if $A\big(\ss(\underline\yy_i), \xxi(\underline\eeta_i)\big) = A\big(\xxi(\underline\eeta_i), \ss(\underline\yy_i)\big)$. \yl{When $m^* = d^* = 1$, coefficients of determination based on regressions (e.g., CTT reliability and PRMSE) are usually asymmetric unless the regressions are linear in both directions.} %\footnote{The case when $m^*>1$ or $d^*>1$ need not be considered because the cofficient of determination is not defined for multidimensional outcomes.}
  %Symmetry is therefore optional. %However, it is valuable when we do not want to decide on the direction of the regression, %Nevertheless, when we aim to emphasize reliability as a property of observed scores or latent scores, more often than not do we care about whether the variability in observed scores or latent scores is to be partitioned and thus need not impose symmetry when choosing among measures of association. \jp{[Sentence doesn't seem essential, and worded in a difficult way to understand.]}
  Symmetry is an optional desideratum which might be desirable in specific contexts. First, symmetry is helpful when it is difficult to unequivocally designate either the observed or latent scores as the regression outcome (e.g., measurement versus prediction decompositions). Second, symmetry can avoid potential confusion between two different valued asymmetric measures of association about the same observed and latent scores, which typically occurs with nonlinear measurement models (e.g., IRT; see \citeNP{liu.pek&maydeu-olivares.2024}). Symmetric measures of association can be formulated using cross-product moments (e.g., the squared or absolute Pearson correlation; the maximal correlation; \citeNP{gebelein.1941}), joint cumulative distribution functions \cite<cdfs; e.g.,>{blum&kiefer&rosenblatt.1961, hoeffding.1948}, ranks \cite<e.g.,>{kruskal.1958}, copulas \cite<e.g.,>{schweizer&wolff.1981}, mutual information and entropy \cite<e.g.,>{joe.1989}, distance metrics between pdfs \cite<e.g.,>{ali&silvey.1965}, and distance covariance \cite<e.g.,>{szekely&rizzo&bakirov.2007}. Readers are referred \citeA{tjostheim&otneim&stove.2022} for a comprehensive review.

\noindent
\textbf{Invariance}\\

Invariance is related to transformations applied to observed and latent scores. Let ${\cal F}$ and ${\cal H}$ be two suitable families of transformations supported on $\real^{m^*}$ and $\real^{d^*}$, respectively. The association measure is invariant with respect to the pair of transformation families $({\cal F}, {\cal H})$ if \textit{$A\big(f(\ss(\underline\yy_i)), h(\xxi(\underline\eeta_i))\big)$ = $A\big(\ss(\underline\yy_i), \xxi(\underline\eeta_i)\big)$ for all $f\in{\cal F}$ and $h\in{\cal H}$}. In words, the association measure remains unchanged under certain transformations of observed and latent scores. The expression above can accommodate potentially different families of transformations (i.e., $\cal F$ and $\cal H$) for the two sets of scores. Observe that regression-based coefficients of determination satisfy a form of invariance. Consider regressing $s(\yy_i)\in\real$ onto $\eeta_i\in\real^{d}$ (i.e., a measurement decomposition). If we set ${\cal F} = \{$all invertible linear transformations on $\real \}$ and ${\cal H} = \{$all invertible transformations on $\real^{d}\}$, then the corresponding coefficient of determination is invariant with respect to $({\cal F}, {\cal H})$.\footnote{\yl{To see why, let the original regression be expressed by $s(\yy_i) = \omega(\eeta_i) + \varepsilon_i$ with fitted value $\omega(\eeta_i) = \ex[s(\underline\yy_i)|\eeta_i]$ and error term $\varepsilon_i$. Take any $f\in{\cal F}$ and $h\in{\cal H}$ such that $f(x) = a + bx$ with $b\ne 0$ and $h$ has a well-defined inverse $h^{-1}$. Then $f(s(\yy_i))$ = $a + bs(\yy_i)$ = $b(\omega\circ h^{-1})(h(\eeta_i))$ + $(a + b\varepsilon_i)$, which can be viewed as a regression onto $h(\eeta_i)$ with predicted value $b(\omega\circ h^{-1})(h(\eeta_i))$ and error term $a + b\varepsilon_i$. This claim follows from observing that $b(\omega\circ h^{-1})(h(\eta_i))$ = $b\omega(\eta_i)$ is uncorrelated with $a + b\epsilon_i$, as implied by the original measurement decomposition of $s({\bf y}_i)$. Because the same linear transform is applied to both the outcome and error, the coefficient of determination remains intact.}} Similarly, when $m^* = d^* = 1$, the squared and absolute Pearson correlation are invariant to invertible linear transformations.

  Coefficients of determination and Pearson correlations, however, are not invariant with respect to nonlinear transformations.  For instance, let $\xi(\underline\eeta_i)\in\real$ follow a standard normal distribution and let $\Phi$ denote its cdf.  Then the percentile rank $\tilde\xi(\eeta_i) = 100\Phi(\xi(\eeta_i))$ is a strictly monotone transformation of the original latent score $\xi(\eeta_i)$. Because of the nonlinearity of $\Phi$, the PRMSEs for predicting $\xi(\eeta_i)$ versus $\tilde\xi(\eeta_i)$ by their respective EAP estimates are often not the same. In contrast, a measure of association satisfying invariance with respect to strictly monotone transformations would yield identical reliability coefficients in both scenarios. Invariance might have intuitive appeal based on the expectation that observed data should carry the same information in predicting related latent quantities that have a one-to-one correspondence.

Several symmetric association measures cited in the ``Symmetry'' section satisfy invariance beyond linear transformations. For asymmetric measures of association, the coefficient considered by \citeA{azadkia&chatterjee.2021}, which generalizes \citeA{chatterjee.2021} and \citeA{dette&siburg&stoimenov.2013}, is invariant to strictly monotone transformations of the outcome and might be used as an alternative to coefficients of determination in measurement and prediction decompositions.% \yl{We present examples of both symmetry and asymmetry in the next section.}

\noindent
\textbf{Examples}\\

\noindent
\textbf{\textit{Absolute and Squared Pearson Correlation}}\\

  We consider first the simplest case in which observed and latent scores are unidimensional (i.e., $m^* = d^* = 1$). \citeA{weiss.1982} computed the (Pearson) correlation (termed a ``fidelity correlation") between true and estimated latent ability scores to evaluate different adaptive testing strategies. Because estimated ability scores usually correlate positively with true ability scores under a unidimensional IRT model, the fidelity correlation can be conceived as a reliability coefficient using the absolute correlation as the association measure. Similarly, \citeA{kim.2012} referred to the squared correlation between a pair of true and estimated latent ability scores as a squared-correlation reliability. For simplicity, we only consider squared correlation below. Let $\underline u$ and $\underline v\in\real$ be two random scalars. The squared correlation between $\underline u$ and $\underline v$ can be expressed as
\begin{equation}
  \corr^2\big(\underline u, \underline v\big) = \frac{\cov(\underline u, \underline v)^2}{\var(\underline u)\var(\underline v)}.
  \label{eq:sqcorr}
\end{equation}
Note that $\corr^2(\underline u, \underline v)$ is distinct from the coefficient of determination $\varrho^2(\underline u, \underline v)$. The two quantities coincide only when the regression of $u$ on $v$ is linear (e.g., when $u$ is a scalar-valued observed score and $v$ is the CTT true score underlying $u$). The squared correlation satisfies the estimability, normalization, and symmetry desiderata, and is only invariant to non-vanishing linear transformations of $u$ and $v$.

\noindent
\textbf{\textit{Coefficient Sigma}}\\

 Let us continue assuming that the observed and latent scores are unidimensional. To allow for nonlinear associations while achieving invariance of nonlinear transformations, symmetric measures of association based on \citeauthor{renyi.1959}'s Axioms can be substituted in place of the absolute or squared correlation. Let
\begin{equation}
  \tilde\varsigma\big(\underline u, \underline v\big) = 4\sin^2\left(\frac{\pi}{6}\varsigma(\underline u, \underline v)\right)
  \label{eq:SW}
\end{equation}
  be the rescaled coefficient sigma,\footnote{In \citeA{schweizer&wolff.1981}, coefficient sigma $\varsigma$ was defined only for continuous random variables. Here, we extend its use to possibly discrete scores. For example, the observed scores are discrete when the MVs are discrete, and some measurement models (e.g., latent class models) incorporate discrete LVs which further results in discrete latent scores. Note that a coefficient sigma computed for discrete scores no longer exactly satisfies the R\'enyi's Axioms.} with
\begin{equation}
  \varsigma\big(\underline u, \underline v\big) = 12\iint_{\real^2}\big|F_{u, v}(s, t) - F_{u}(s)F_{v}(t)\big|F_{u}(ds)F_{v}(dt)
  \label{eq:OSW}
\end{equation}
as the original coefficient sigma \cite{schweizer&wolff.1981}. In Equation \ref{eq:OSW}, $F_{u,v}$ denotes the joint cdf of $\underline u$ and $\underline v$, and $F_{u}$ and $F_{v}$ are the marginal cdfs of $\underline u$ and  $\underline v$, respectively. The original coefficient sigma, $\varsigma$ (Equation \ref{eq:OSW}) then measures the average absolute deviation between the actual joint distribution of two scores, $F_{u, v}(s, t)$, and the simpler joint distribution in which $\underline u$ is independent of $\underline v$, $F_{u}(s)F_{v}(t)$. Equation \ref{eq:OSW} is a successive integral over the marginal distributions of $\underline u$ and $\underline v$ and the integrand only depends on the cdfs; thus, $\varsigma(\underline u, \underline v)$ is invariant to one-to-one transformations of $\underline u$ and $\underline v$. The transformation in Equation \ref{eq:SW} is monotone, guaranteeing that $\tilde\varsigma(\underline u, \underline v)$ coincides with the squared Pearson correlation when $\underline u$ and $\underline v$ follow a bivariate normal distribution \cite{schweizer&wolff.1981}. The original coefficient sigma, $\varsigma$, is also closely related to Spearman's correlation, which is obtained by replacing the absolute difference in Equation \ref{eq:OSW} by the signed difference. If the two scores are positively quadrant dependent,\footnote{$\underline u$ and $\underline v$ are positive quadrant dependent if $F_{u,v}(s, t) \ge F_{u}(s)F_{v}(t)$ for all $s,t\in\real$ \cite[Definition 5.2.1]{nelsen.2006}.} then the original coefficient sigma, $\varsigma$, and Spearman's correlation are identical \cite[p. 209]{nelsen.2006}. The rescaled coefficient sigma (Equation \ref{eq:SW}) satisfies estimability, normalization, and symmetry; it is also invariant to strictly monotone transformations for both $\underline u$ and $\underline v$.

\noindent
\textbf{\textit{Mutual Information}}\\
  This example illustrates a symmetric association measure when both the observed and latent scores are potentially multidimensional. The mutual information between random vectors $\underline\uu$ and $\underline\vv$ of any dimension can be expressed as
\begin{equation}
  M\big(\underline\uu, \underline\vv\big) = \iint \log\left[\frac{f_{\uu, \vv}(\ss, \tt)}{f_{\uu}(\ss)f_{\vv}(\tt)}\right]F_{\uu, \vv}(d\ss, d\tt),
  \label{eq:mi}
\end{equation}
in which $f_{\uu, \vv}$ denotes the joint pdf of $\underline\uu$ and $\underline\vv$, $f_{\uu}$ denotes the marginal pdf of $\underline\uu$, and $f_{\vv}$ denotes the marginal pdf of $\underline\vv$. Mutual information (Equation \ref{eq:mi}) is the Kullback-Leibler divergence of the true joint pdf of $\underline\uu$ and $\underline\vv$, $f_{\uu, \vv}(\ss, \tt)$, from the simpler pdf in which the two random vectors are independent, $f_{\uu}(\ss)f_{\vv}(\tt)$. Thus, mutual information is non-negative and attains zero if and only if $\underline\uu$ and $\underline\vv$ are independent. From Equation \ref{eq:mi}, mutual information is also symmetric and invariant to invertible transformations of $\underline\uu$ and $\underline\vv$. However, mutual information is not bounded from above. To normalize mutual information to the unit interval, \citeauthor{joe.1989} (\citeyearNP{joe.1989}; see also \citeNP{linfoot.1957}) proposed rescaling $M$ by
\begin{equation}
  \tilde M(\underline\uu, \underline\vv) = 1 - \exp\left[-2M(\underline\uu, \underline\vv)\right].
  \label{eq:MI}
\end{equation}
When $\underline\uu$ and $\underline\vv$ follow jointly a multivariate normal distribution, \citeA{joe.1989} showed that $\tilde M$ reduces to the squared Pearson correlation when both random vectors reduce to random scalars (i.e., $\underline\uu = \underline u$ and $\underline\vv = \underline v$); $\tilde M$ also reduces to the coefficient of determination when one of the two random quantities is unidimensional and used as the regression outcome. These special cases justify the normalization of mutual information by mapping $x\mapsto1 - \exp(-2x)$. %Alternative normalizing options for mutual information can be found in \citeA{joe.1989}.
Mutual information has been applied to quantify measurement precision in measurement models with both discrete and continuous LVs \cite<e.g.,>{chen&liu&xu.2018, johnson&sinharay.2020, markon.2013, markon.2023, sinharay&johnson.2019}. The rescaled mutual information (Equation \ref{eq:MI}) satisfies the estimability, normalization, and symmetry desiderata, and is invariant to invertible transformations of $\underline\uu$ and $\underline\vv$.

  \noindent
  \textbf{\textit{\boldmath Coefficient $T$}}\\
  The third example features an asymmetric measure, in which we find an alternative to the coefficient of determination that is invariant to strictly monotone transformations of the outcome variable. Let $\underline u\in\real$ be a scalar outcome variable and $\underline\vv$ be a set of explanatory variables. Define the Azadkia-Chatterjee coefficient $T$ as
\begin{equation}
  T(\underline u, \underline\vv) = \frac{\int_\real\var\big(\pr\{\underline u>s|\underline\vv\}\big)F_{u}(ds)}{\int_\real\var\big(\ind\{\underline u>s\}\big)F_{u}(ds)},
  \label{eq:AC}
\end{equation}
in which $\pr\{\underline u > s|\vv\}$ denotes the conditional probability of $\underline u > s$ given $\vv$ and $\ind\{\underline u > s\}$ is the indicator function of when $\underline u > s$. Equation \ref{eq:AC} also pertains to a signal-to-total ratio \cite<STR; cf.,>{cronbach&gleser.1964}, analogous to the coefficient of determination. Recall that a coefficient of determination quantifies the amount of variance in the outcome (i.e., total information) that is taken into account by the predictor variables (i.e., signal) on the normalized scale. In a similar vein, the coefficient $T$ partitions the total variability of a threshold-passing indicator of the outcome $\ind\{\underline u>s\}$ and reflects the portion of the systematic variation ascribed to the predictor variables $\underline \vv$, omitting the leftover variance unassociated with $\underline \vv$. Because the threshold  $s$ is arbitrarily chosen, the systematic and total variability are then respectively integrated across all possible values of $s$ under the marginal distribution of $\underline u$. Invariance to strictly monotone transformations of the outcome variable follows from the use of the indicator function as well as the integral with respect to the outcome distribution. Similar to coefficients of determination, coefficient $T$ is only applicable in the regression framework (i.e., $T$ is asymmetric) and is estimable, normalized, and invariant to invertible transformations of explanatory variables. In addition, coefficient $T$ is invariant to strict monotone transformations to the outcome whereas a coefficient of determination is only invariant to non-vanishing linear transformations.
%As an example, consider the measurement decomposition whereby a unidimensional observed score $s(\underline\yy_i)$ is predicted by LVs in $\underline\eeta_i$. When invariance to a monotonically transformed observed score is desired, let reliability be defined by $T\big(s(\underline\yy_i), \underline\eeta_i\big)$ instead of the coefficient of determination $\varrho^2\big(s(\underline\yy_i), \underline\eeta_i\big)$. Alternatively, in a prediction decomposition whereby a unidimensional latent score $\xi(\underline\eeta_i)$ is predicted by MVs $\underline\yy_i$, we can substitute $\varrho^2\big(\xi(\underline\eeta_i), \underline\yy_i\big)$ with $T\big(\xi(\underline\eeta_i), \underline\yy_i)\big)$ because the latter is invariant to monotone transformations of the latent score.

\noindent
\textbf{\textit{\yl{Generalized Coefficient of Determination}}}\\
  This example illustrates how measurement and prediction decompositions can be generalized to allow for multiple outcomes and free choice of explanatory variables. Given observed scores $\ss(\yy_i)\in\real^{m^*}$ and latent scores $\xxi(\eeta_i)\in\real^{d^*}$. Let a \textit{generalized measurement decomposition} be defined by
\begin{equation}
  \ss(\yy_i) = \ex\big[\ss(\underline\yy_i)|\xxi(\eeta_i)\big] + \boldsymbol\varepsilon_i^*,
  \label{eq:genmeasdecomp}
\end{equation}
and a \textit{generalized prediction decomposition} be defined by
\begin{equation}
  \xxi(\eeta_i) = \ex\big[ \xxi(\underline\eeta_i)|\ss(\yy_i) \big] + \boldsymbol\delta_i^*.
  \label{eq:genpreddecomp}
\end{equation}
In Equations \ref{eq:genmeasdecomp} and \ref{eq:genpreddecomp}, their outcome variables (i.e., $\ss(\yy_i)$ and $\xxi(\eeta_i)$)  and corresponding error terms (i.e., $\boldsymbol\varepsilon_i^*$ and $\boldsymbol\delta_i^*$) can be multidimensional (cf. Equations \ref{eq:measdecomp} and \ref{eq:preddecomp} for measurement and prediction decompositions, respectively). Moreover, the explanatory variables that are being conditioned on the right-hand side of Equations \ref{eq:genmeasdecomp} and \ref{eq:genpreddecomp} can be any latent scores (cf. only LVs or CTT true scores in Equation \ref{eq:measdecomp}) and any observed score (cf. only MVs or EAP scores in Equation \ref{eq:preddecomp}), respectively. Various coefficients quantifying STR can be computed for multivariate regression models, generalizing the coefficient of determination.

  Let $\underline\uu$ and $\underline\vv$ be multiple outcome and explanatory variables, respectively. Then, the multivariate regression of $\underline\uu$ on $\underline\vv$ is
\begin{equation}
  \uu = \ex\big(\underline\uu|\vv\big) + \ee,
  \label{eq:mreg}
\end{equation}
which subsumes Equations \ref{eq:genmeasdecomp} and \ref{eq:genpreddecomp} as special cases. The error vector in Equation \ref{eq:mreg}, $\ee$, satisfies $\cov(\underline\ee)$ = $\cov(\underline\uu)$ $-$ $\cov\big[\ex(\underline\uu|\vv)\big]$, which is the multivariate analog to the law of total variance. \yl{A generalization for coefficients of determination in multivariate regression (Equation \ref{eq:mreg}) is:}

\begin{equation}
 W(\underline\uu, \underline\vv) = 1 - \frac{\det\big(\cov(\underline\ee)\big)}{\det\big(\cov(\underline\uu)\big)} = \frac{\det\big(\cov(\underline\uu)\big) - \det\big(\cov(\underline\ee)\big)}{\det\big(\cov(\underline\uu)\big)}.
 \label{eq:wilks}
\end{equation}
\yl{Coefficient $W$ (Equation \ref{eq:wilks}) is an population counterpart of (one minus) Wilks' lambda in multivariate regression \cite{wilks.1932}}, in which noise is quantified by the error covariance matrices $\cov(\underline\ee)$ and signal is quantified by the total covariance matrix $\cov(\underline\uu)$ minus the error covariance matrix. The matrix determinant, $\det(\cdot)$, is taken to obtain a single-number summary of covariance matrices, which \citeA{wilks.1932} referred to as the generalized variance. It can be verified that Equation \ref{eq:wilks} reduces to the coefficient of determination $\varrho^2(\underline u, \underline\vv)$ when the outcome variable $u$ is unidimensional. \yl{Alternative multivariate STR measures can be constructed from, for instance, Pillai's trace and Roy's largest root \cite<e.g.,>{mardia&kent&bibby.1979}, which are not further considered here due to limited space. Coefficient $W$ (Equation \ref{eq:wilks}) is estimable, normalized, but not symmetric;} they are invariant to invertible transformations of explanatory variables and non-vanishing linear transformations of outcome variables.

\centerline{\textbf{Numerical Study}}

We conducted a numerical study to illustrate the behavior of various reliability coefficients at the level of the population. We examined (a) how the numerical values of these reliability measures change as functions of test length under a two-dimensional simple-structure IRT model, and (b) how they map onto other benchmarks of measurement error (e.g., estimation error of latent scores and inter-LV correlations).

\noindent \textbf{\textit{Data Generation}}\\
\begin{figure}[!t]
\centering
\begin{tikzpicture}
\tikzset{lv/.style={draw,circle,thick,inner sep=2pt,minimum size=35pt}};
\tikzset{mv/.style={draw,rectangle,thick,inner sep=2pt,minimum width=40pt,minimum height=30pt}};
\tikzset{arrow1/.style={-latex,thick}};
\tikzset{arrow2/.style={latex-latex,thick}};
\tikzset{par/.style={draw=none,fill=white,font=\footnotesize,inner sep=0pt}};
% first LV
\node[style=lv] at (0, 0) (eta1) {$\eta_{i1}$};
\node[draw=none,above=50pt of eta1]  (c1) {};
\node[draw=none, right=0pt of c1]  (d1) {$\cdots$};
\node[style=mv, left=0pt of c1]  (y2) {$y_{i2}$};
\node[style=mv, left=8pt of y2]  (y1) {$y_{i1}$};
\node[style=mv, right=8pt of d1]  (ymh) {$y_{i,\frac{m}{2}}$};
\draw[style=arrow1] (eta1) -- (y1.south);
\draw[style=arrow1] (eta1) -- (y2.south);
\draw[style=arrow1] (eta1) -- (ymh.south);
\draw[style=arrow2] ([shift=(-50:-1pt)]eta1.south west)
  arc[start angle=160,end angle=-120,radius=-10pt];
% second LV
\node[style=lv, right=150pt of eta1] (eta2) {$\eta_{i2}$};
\node[draw=none,above=50pt of eta2]  (c2) {};
\node[draw=none, right=0pt of c2]  (d2) {$\cdots$};
\node[style=mv, left=0pt of c2]  (ymhp2) {$y_{i,\frac{m}{2}+2}$};
\node[style=mv, left=8pt of ymhp2]  (ymhp1) {$y_{i,\frac{m}{2}+1}$};
\node[style=mv, right=8pt of d2]  (ym) {$y_{im}$};
\draw[style=arrow1] (eta2) -- (ymhp1.south);
\draw[style=arrow1] (eta2) -- (ymhp2.south);
\draw[style=arrow1] (eta2) -- (ym.south);
\draw[style=arrow2] ([shift=(70:14pt)]eta2.south east)
  arc[start angle=120,end angle=-160,radius=10pt];
  \draw[style=arrow2] (eta1.south east) .. controls (1.5, -1.5) and (5, -1.5).. %
  (eta2.south west);
\end{tikzpicture}
\caption{Path diagram for the two-dimensional measurement model. $\eta_{i1}$ and $\eta_{i2}$ are latent variables and $y_{i1}, \dots, y_{im}$ are manifest variables.}
  \label{fig:path}
\end{figure}

  Figure \ref{fig:path} presents the data generating model in which the total number of MVs $m$ (i.e., test length) is even such that each LV is indicated by the same number of MVs. The two LVs follow a  bivariate normal distribution:
\begin{equation}
  \underline\eeta_i = (\underline\eta_{i1}, \underline\eta_{i2})\t\sim{\cal N}\left(
    \begin{bmatrix}
      0\\0
    \end{bmatrix},
    \begin{bmatrix}
      1 & 0.5\\
      0.5 & 1
    \end{bmatrix}
  \right).
  \label{eq:lvdist}
\end{equation}
  Conditional on $\eeta_i$, every MV is mutually independent of one another (i.e., local independence). Each MV $y_{ij}\in\{0, 1\}$ and the conditional probability of $\underline y_{ij} = 1$ given $\eeta_i$ follows a three-parameter logistic model (\citeNP{birnbaum.1968}):
\begin{equation}
  \pr\{\underline y_{ij} = 1|\eeta_i\} = c_j + \frac{1 - c_j}{1 + \exp\left[  -a_j(\eta_{i,k(j)} - b_j)\right]},
  \label{eq:3pl}
\end{equation}
in which $a_j$, $b_j$, and $c_j$ are the discrimination, difficulty and pseudo-guessing parameters, respectively. Furthermore, $j = 1,\dots, m$ indexes the MVs, and $k(j) = 1$ if $j\le m/2$ and 2 otherwise. We varied the test length from $m = 6$ to 120 at increasing intervals of 6. For each level of $m$, item parameters were independently drawn from the following distributions: $\underline a_j\sim\hbox{Uniform}(0.5, 2)$, $\underline b_j\sim\hbox{Uniform}(-2, 2)$, and $\underline c_j\sim\hbox{Uniform}(0, 0.2)$, $j = 1,\dots, m$. For each unique set of item parameters (i.e., replication), we generated 1000 MC samples of LV and MV vectors from which we estimated reliability coefficients and benchmark measures.

\noindent \textbf{\textit{Scores, Reliability Measures, and Benchmarks}}\\

Two pairs of observed and latent scores were considered in the simulation. First, we are interested in estimating the LV score $\eeta_i = (\eta_{i1}, \eta_{i2})\t$ by the corresponding EAP score $\ex(\underline\eeta_i|\yy_i)$. For reliability measures that can handle multivariate scores, let $\ss(\yy_i) = \ex(\underline\eeta_i|\yy_i)$ and $\xxi(\eeta_i) = \eeta_i$. Only the first element of a two-dimensional score vector is considered if the reliability measure only applies to unidimensional scores; i.e., $s_1(\yy_i) = \ex(\underline\eta_{i1}|\yy_i)$ and $\xi_1(\eeta_i) = \eta_{i1}$. Second, to illustrate the impact of monotone transformations, we used the same observed scores but transformed the latent scores into their percentile ranks, resulting in $\ss(\yy_i) = \ex(\underline\eeta_i|\yy_i)$ and $\xxi(\eeta_i) = \left(100\Phi(\eta_{i1}), 100\Phi(\eta_{i2})\right)\t$. Whenever a unidimensional score is required, we \yl{specify} $s_1(\yy_i) = \ex(\underline\eta_{i1}|\yy_i)$ and $\xi_1(\yy_i) = 100\Phi(\eta_{i1})$.

\begin{table}[!t]
  %\color{blue1}
  \centering
  \caption{Summary of various reliability coefficients based on pairs of observed and latent scores, symmetry about the two scores, and invariance under the percentile-rank transform of latent scores. \yl{Asterisks (*) are added to indicate novel reliability coefficients that have not been considered in the reliability literature.}  Measure = observed scores as outcome, predict = latent scores as outcome, $\varrho^2$ = Coefficient of determination, $\corr^2$ = squared Pearson correlation, Sigma = \yl{rescaled} coefficient sigma (Equation \ref{eq:SW}), $T$ = coefficient $T$ (Equation \ref{eq:AC}), MI = rescaled mutual information (Equation \ref{eq:MI}), and \yl{$W$ = coefficient $W$ (Equation \ref{eq:wilks})}, $s_1(\yy_i)$ = unidimensional observed score, $\xi_1(\eeta_i)$ = unidimensional latent score, $\ss({\yy_i})$ = two-dimensional observed scores, and $\xxi(\eeta_i)$ = two-dimensional latent scores.
  }
  \label{tab:reliab}
  \begin{tabular}{ccccc}
    \toprule
   Coefficient & Observed & Latent & Symmetry & Invariance\\
    \midrule
    $\varrho^2$ (measure) & $s_1(\yy_i)$ & $\xxi(\eeta_i)$ & no & yes\\
    $\varrho^2$ (predict) & $\ss(\yy_i)$ & $\xi_1(\eeta_i)$ & no & no \\
   $\corr^2$ & $s_1(\yy_i)$ & $\xi_1(\eeta_i)$ & yes & no \\
   \color{blue1}Sigma$^*$ & $s_1(\yy_i)$ & $\xi_1(\eeta_i)$ & yes & yes \\
   \color{blue1}$T$ (measure)$^*$ & $s_1(\yy_i)$ & $\xxi(\eeta_i)$ & no & yes\\
   \color{blue1}$T$ (predict)$^*$ & $\ss(\yy_i)$ & $\xi_1(\eeta_i)$ & no & yes \\
    MI & $\ss(\yy_i)$ & $\xxi(\eeta_i)$ & yes & yes \\
    \color{blue1}$W$ (measure)$^*$ & $\ss(\yy_i)$ & $\xxi(\eeta_i)$ & no & yes \\
    \color{blue1}$W$ (predict)$^*$ & $\ss(\yy_i)$ & $\xxi(\eeta_i)$ & no & no \\
    \bottomrule
  \end{tabular}
\end{table}

Nine reliability association measures were investigated. Table \ref{tab:reliab} provides a summary of the association measures, observed scores, and latent scores involved in each coefficient, as well as whether or not the coefficient is symmetric and invariant to the percentile-rank transformation of latent scores. When the latent scores are the original LVs (i.e., $\xxi(\eeta_i) = \eeta_i$), observe that (a) the coefficient of determination for the regression of $s_1(\yy_i)$ onto $\xxi(\eeta_i)$ coincides with the CTT reliability of $s_1(\yy_i)$, and that (b) the coefficient of determination for the regression of $\xi_1(\eeta_i)$ onto $\ss(\yy_i)$ is identical to the squared correlation between $s_1(\yy_i)$ and $\xi_1(\eeta_i)$, which further equals to PRMSE of $\xi_1(\eeta_i)$.

\begin{figure}[!t]
  \centering
  \includegraphics[width=\textwidth]{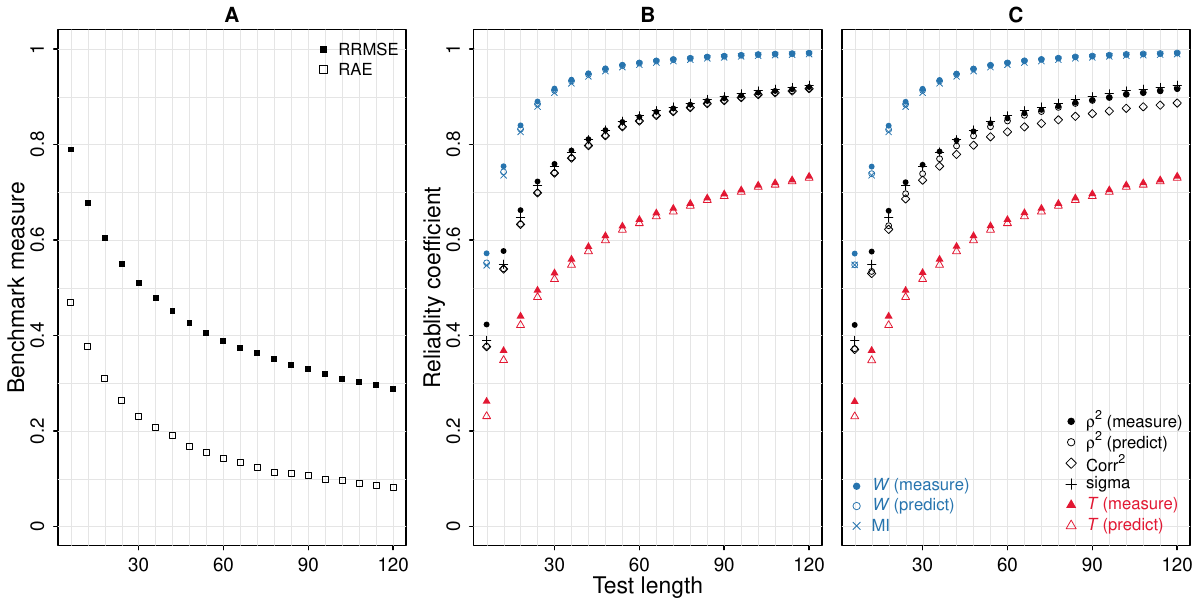}
  \caption{Two benchmark measures (panel A) and relability measures (panels B and C) as functions of test length. \yl{Panel B summarizes results when latent scores are original LV scores, and panel C summarizes results when latent scores are precentile ranks of LV scores.} RRMSE = root relative mean squared error in latent variable scores, RAE = relative absolute error in inter-latent-variable correlation,  measure = observed score as outcome, predict = latent score as outcome, $\varrho^2$ = coefficient of determination ($\varrho^2$ (measure) = CTT reliability and $\varrho^2$(predict = PRMSE), $\corr^2$ = squared Pearson correlation, sigma = rescaled coefficient sigma (Equation \ref{eq:SW}), $T$ = coefficient $T$ (Equation \ref{eq:AC}), MI = rescaled mutual information (Equation \ref{eq:MI}), and \yl{$W$ = coefficient $W$ (Equation \ref{eq:wilks}).}}
  \label{fig:sim}
\end{figure}

\yl{
  Within each replication, we estimated all the reliability coefficients empirically based on 1000 MC samples using the procedure introduced in \citeA{liu.pek&maydeu-olivares.2024}. A brief summary of the procedure is included in the Supplementary Materials. EAP scores were computed using the R package \texttt{mirt} \cite{chalmers.2012} with item parameters fixed at the data generating values. To estimate coefficients of determination and $W$, we obtained predicted values and residuals by nonparametric regression. In particular, we applied the default thin-plate spline smoother from the \texttt{mgcv} package \cite{wood.2003}. Note that the numerical results are not sensitive to the choice of nonparametric regressors. As evidence, we reproduced the results in Figure \ref{fig:sim} using local polynomial regression \cite<by the R function \texttt{loess};>{cleveland.1979, cleveland.et.al.2017} instead of regression splines; these additional results are reported in the Supplementary Materials. We estimated coefficient sigma with empirical copulas \cite[Section 5.6]{nelsen.2006} using the \texttt{wolfCOP} function in the \texttt{copBasic} package \cite{asquith.2023}. Mutual information was estimated using a method based on nearest neighbor distances \cite{kraskov.et.al.2004}, which was implemented in the \texttt{knn\_mi} function from the \texttt{rmi} package \cite{michaud.2018}. The coefficient $T$ can be empirically estimated by the CODEC coefficient $T_n$ \cite[p. 3072]{azadkia&chatterjee.2021}, which we computed by calling the \texttt{codec} function in the \texttt{FOCI} package \cite{azadkia&chatterjee&matloff.2021}. To aid in the accessibility of our developments, example R code is provided in the Supplementary Materials.
}

Two additional benchmark measures were computed to reflect the recovery of LV scores $\eeta_i$ and the inter-LV correlation relative to the sizes of true values. The root relative mean squared error (RRMSE) is defined as
\begin{equation}
  \mathrm{RRMSE} = \sqrt{\frac{\sum_{i=1}^{1000}\sum_{k=1}^2\left(\ex(\underline\eta_{ik}|\yy_i) - \eta_{ik}\right)^2}{\sum_{i=1}^{1000}\sum_{k=1}^2\eta_{ik}^2}}
  \label{eq:rmse},
\end{equation}
in which $i$ indexes each MC draw and $k = 1, 2$ indexes the dimensions of LVs. RRMSE measures the overall estimation error of $\eeta_i$ by their EAP scores $\ex(\underline\eeta_i|\yy_i)$. The relative absolute error (RAE) reflects how well the correlation between EAP scores approximates the true inter-LV correlation (0.5; see Equation \ref{eq:lvdist}):
\begin{equation}
  \mathrm{RAE} = \frac{|\widehat{\corr}\big(\ex(\underline\eta_{i1}|\yy_i), \ex(\underline\eta_{i2}|\yy_i)\big) - 0.5|}{0.5},
  \label{eq:rae}
\end{equation}
in which $\widehat{\corr}$ denotes the empirical Pearson correlation computed from 1000 MC draws. Values from Equations \ref{eq:rmse} and \ref{eq:rae} are expected to decrease as the test length $m$ grows because increasing $m$ is associated with more consistent estimates of EAP scores.

\noindent \textbf{\textit{Results}}\\

\yl{We averaged various benchmark measures and reliability coefficients across multiple sets of item parameters and present them as functions of test length in Figure \ref{fig:sim}.} With increasing test length $m$, the two benchmark measures of estimation error (PRMSE and RAE) monotonically decrease (see Figure \ref{fig:sim}A), indicating better recovery of LV scores and inter-LV correlations. Although we place PRMSE and RAE within the same plot in which numerical values fall within the unit interval, these values are not directly comparable because they quantify different aspects of the estimates. The RAE is a scalar-valued measure about the inter-LV correlation (ranging from .08 to .47) and RRMSE is a measure for multiple random quantities (i.e., two-dimensional LVs; ranging from .29 to .79).

In Figure \ref{fig:sim}B, reliability coefficients increase in value as test length $m$ increases, indicating that EAP scores become better proxies of LVs. Different association measures are not always comparable even though they have been normalized because different reliability coefficients are defined for potentially different pairs of observed and latent scores while quantifying distinct forms of association (see Table \ref{tab:reliab}). Figure \ref{fig:sim}B suggests that the nine reliability coefficients cluster into three groups (shown in different colors). Coefficients of determination (corresponding to CTT reliabliity and PRMSE) together with the rescaled coefficient sigma, are very similar in value across all levels of $m$ (approximately from 0.4 to 0.9). The squared correlation coincides with PRMSE in the population; hence, \yl{the estimated squared correlation} and PRMSE exhibit almost identical values in the simulation. CTT reliability is observed to be at least as large as PRMSE, which is a known result \cite[Equation 31]{kim.2012}. Rescaled sigma lies between CTT reliability and PRMSE when $m$ is small and becomes the largest among the three reliability indexes when $m$ is large. The measurement and prediction versions of coefficient $T$ take on smaller values compared to the coefficients of determination and rescaled sigma. Coefficient $T$ for the measurement decomposition (ranging from .26 to .73) is slightly larger than the coefficient for the prediction decomposition (ranging from .23 to .73), especially at smaller $m$. Finally, the three association measures between the two-dimensional LVs and the two-dimensional EAP scores are the largest in magnitude at all levels of $m$ (approximately ranging .55 and .99). \yl{Coefficient $W$s under generalized measurement decompositions are uniformly larger than those from generalized prediction decompositions, which are in turn uniformly larger than rescaled mutual information.}

Transforming LVs to their percentile ranks leaves most coefficients under investigation intact. However, transforming the LVs changes the squared correlation, coefficient of determination based on the prediction decomposition of $\xi_1(\eeta)$, and coefficient $W$ based on the generalized prediction decomposition of $\xxi(\eeta)$. In Figure \ref{fig:sim}C, the squared correlation and the prediction $\varrho^2$ are lower than their values in Figure \ref{fig:sim}B; moreover, the transformation destroys the equivalence between the two coefficients. \yl{Coefficient $W$s under generalized prediction decompositions were observed to decrease slightly because of the LV transformation (see Figure \ref{fig:sim}B versus \ref{fig:sim}C).}

  \centerline{\textbf{Summary and Discussion}}

  Reliability is a measure of how closely observed and latent scores align with one another. Based on the regression framework of reliability \cite{liu.pek&maydeu-olivares.2024, mcdonald.2011}, which assumes a LV measurement model, we have shown that reliability can be broadly defined as a measure of association between observed and latent scores (Equation \ref{eq:genrel}). This broad definition subsumes popular indices of reliability that are coefficients of determination such as CTT reliability \cite{lord&novick.1968} and PRMSE \cite{haberman&sinharay.2010}. Because this broad definition of reliability includes very many reliability indices, we identified and described four desiderata that might aid the analyst in selecting the best reliability coefficient(s) for their research. We consider the desiderata of estimability and normalization essential for interpretation. The desiderata of symmetry and invariance, however, are optional depending on the research context.

  From our numerical illustration, we show that different reliability coefficients can be computed from a single measurement model. In general, values of these reliability coefficients increase as a function of test length. Furthermore, association measures between multiple outcome and explanatory variables (e.g., mutual information and coefficient $W$) tend to have larger values compared to association measures based on univariate regression (e.g., CTT reliability and PRMSE). Importantly, these values of reliability cannot be compared with one another despite being normalized onto $[0,1]$, because they measure qualitatively distinct associations between latent and observed scores. %Thus, we recommend that analysts pick the reliability coefficient consistent with their research question, convention, and our desiderata.

 Our general framework expands the notion of reliability in several ways. First, the analyst is not constrained by the choice of observed scores and latent scores to include in a regression. Second, the analyst can choose association measures other than the coefficient of determination. Third, the analyst might move from a univariate regression model (e.g., CTT reliability and PRMSE) to a multivariate regression model (e.g., coefficient $W$). Fourth, reliability coefficients can further be chosen based on symmetry and transformation invariance. Because some reliability coefficients we have described are relatively unfamiliar, future research should study their performance in real-data and simulation settings (e.g., under different LV measurement models). Furthermore, to encourage the application of these novel reliability coefficients by substantive researchers, methodologists would need to develop benchmarks or recommendations on how these distinct measures of reliability might be qualitatively interpreted. It is our hope that this general framework might motivate the development of novel reliability coefficients that are useful to substantive researchers, which have yet to be incorporated in the current work.

 \bibliography{newRel.rev}

 

\section*{Supplementary material}

**Supplementary Materials for**

**"On a General Theoretical Framework of Reliability"**

Yang Liu[1]          Jolynn Pek[2]          Alberto Maydeu-Olivares[3,4]

## Contents

---

[1]Department of Human Development and Quantitative Methodology, University of Maryland, College Park.  Correspondence author.  Email: yliu87@umd.edu

[2]Department of Psychology, the Ohio State University.

[3]Department of Psychology, University of South Carolina

[4]Faculty of Psychology, University of Barcelona

# A Rényi's Axioms

Rényi's Axioms comprise of a set of formal rules that define coefficients of dependence—a specific class of association measures. Because the original Axioms (Rényi, 1959) were too restrictive to be practically useful, several variants have been proposed in the literature (e.g., Geenens & Lafaye de Micheaux, 2022; Nelsen, 2006, p. 208; Schweizer & Wolff, 1981). Here, we present a version of Rényi's Axioms adapted from Schweizer and Wolff (1981).

Let $\underline{u}, \underline{v} \in \mathbb{R}$ be two scalar-valued, continuous random variables. Note that we use an underbar to highlight that the variable is random instead of fixed. A *measure of dependence* for $\underline{u}$ and $\underline{v}$, denoted $A(\underline{u}, \underline{v})$, satisfies the following properties:

*(R1)* $A(\underline{u}, \underline{v})$ *is defined for every pair of random variables $\underline{u}$ and $\underline{v}$.*

*(R2)* $A(\underline{u}, \underline{v}) = A(\underline{v}, \underline{u})$.

*(R3)* $0 \leq A(\underline{u}, \underline{v}) \leq 1$.

*(R4)* $A(\underline{u}, \underline{v}) = 0$ *if and only if $\underline{u}$ and $\underline{v}$ are statistically independent.*

*(R5)* $A(\underline{u}, \underline{v}) = 1$ *if and only if $\underline{u}$ and $\underline{v}$ are almost surely strictly monotone functions of one another.*

*(R6)* *If $f$ and $h$ are almost surely strictly monotone functions on the ranges of $\underline{u}$ and $\underline{v}$, respectively, then $A(f(\underline{u}), h(\underline{v})) = A(\underline{u}, \underline{v})$.*

*(R7)* *If $\underline{u}$ and $\underline{v}$ follow jointly a bivariate normal distribution, then $A(\underline{u}, \underline{v})$ is a strictly increasing function of the squared Pearson correlation.*

*(R8)* *If a sequence of random vectors $(\underline{u}_1, \underline{v}_1)', (\underline{u}_2, \underline{v}_2)', \ldots$ converge in distribution to $(\underline{u}, \underline{v})'$, then $\lim_{n \to \infty} A(\underline{u}_n, \underline{v}_n) = A(\underline{u}, \underline{v})$.*

Despite being more restrictive, Rényi's Axioms motivated our four desiderata for reliability coefficients. (R1) is a global existence condition, which we assume as a prerequisite in our formulation of reliability (see Footnote 5 in the main article). (R8) ensures consistent estimation of the coefficient, which is part of our estimability desideratum. Our requirement of normalization

1

is a combination of (R3)–(R5); (R3) requires the measure to take values on the unit interval, while (R4) and (R5) respectively defines zero and perfect associations. Meanwhile, (R7) is an additional requirement that further enhances interpretability. In particular, (R7) forces the coefficient to be isomorphic to the squared correlation under the familiar scenario of bivariate normality. As for our two optional desiderata, (R2) corresponds to symmetry and (R6) corresponds to invariance with respect to (almost surely) strictly monotonic functions.

# B  Estimating Reliability Coefficients by Monte Carlo

We describe the Monte Carlo procedure that can estimate various reliability coefficients in our numerical study. Our description is slightly more general than Liu, Pek, and Maydeu-Olivares (2024) because we no longer restrict ourselves to the regression framework.

The procedure begins with a known measurement model that specifies the joint distribution of the latent variables (LVs) $\underline{\boldsymbol{\eta}}_i$ and the manifest variables (MVs) $\underline{\mathbf{y}}_i$. In particular, we assume that we can generate (a) $\underline{\boldsymbol{\eta}}_i$ marginally and (b) $\underline{\mathbf{y}}_i | \boldsymbol{\eta}_i$ for almost surely all $\boldsymbol{\eta}_i$. As usual, let $s(\mathbf{y}_i)$ and $\boldsymbol{\xi}(\boldsymbol{\eta}_i)$ denote observed scores and latent scores, respectively. The Monte Carlo procedure follows three steps:

Step 1. *Simulate latent scores.* Generate a large independent sample of LV vectors $\boldsymbol{\eta}_i$, $i = 1, \ldots, M$, in which $M$ denotes the Monte Carlo sample size. Compute latent scores $\boldsymbol{\xi}(\boldsymbol{\eta}_1), \ldots, \boldsymbol{\xi}(\boldsymbol{\eta}_M)$. In general, $M$ should be large to limit Monte Carlo error.

Step 2. *Simulate observed scores.* For each case $i$, simulate an MV vector $\mathbf{y}_i$ conditional on $\boldsymbol{\eta}_i$. Compute observed scores $s(\mathbf{y}_1), \ldots, s(\mathbf{y}_M)$.

Step 3. *Estimate reliability coefficient.* Evaluate an empirical estimator of the association measure $A(s(\underline{\mathbf{y}}_i), \boldsymbol{\xi}(\underline{\boldsymbol{\eta}}_i))$ using the paired Monte Carlo samples $(s(\mathbf{y}_i)', \boldsymbol{\xi}(\boldsymbol{\eta}_i)')'$, $i = 1, \ldots, M$. The resulting estimate is accurate as long as the Monte Carlo sample size $M$ is sufficiently large.

Please refer to the attached R code for an example implementation of the Monte Carlo procedure.

## C    Additional Numerical Results

We examine the extent to which regression-based reliability estimates are sensitive to the choice of nonparametric regression estimators. We re-conducted our numerical study using local polynomial regression (with the R function `loess`; Cleveland, Grosse, & Shyu, 2017) to estimate coefficients of determination and coefficients $W$ (Equation 15). Comparisons between the new results and those reported in the main article, which used the R function `gam` in the `mgcv` package (Wood, 2003; Figure 2), is summarized in Figure S1 below. We conclude that thin-plate spline regression and local polynomial regression yield almost identical reliability estimates across all test length conditions for LVs on both original and transformed scales.

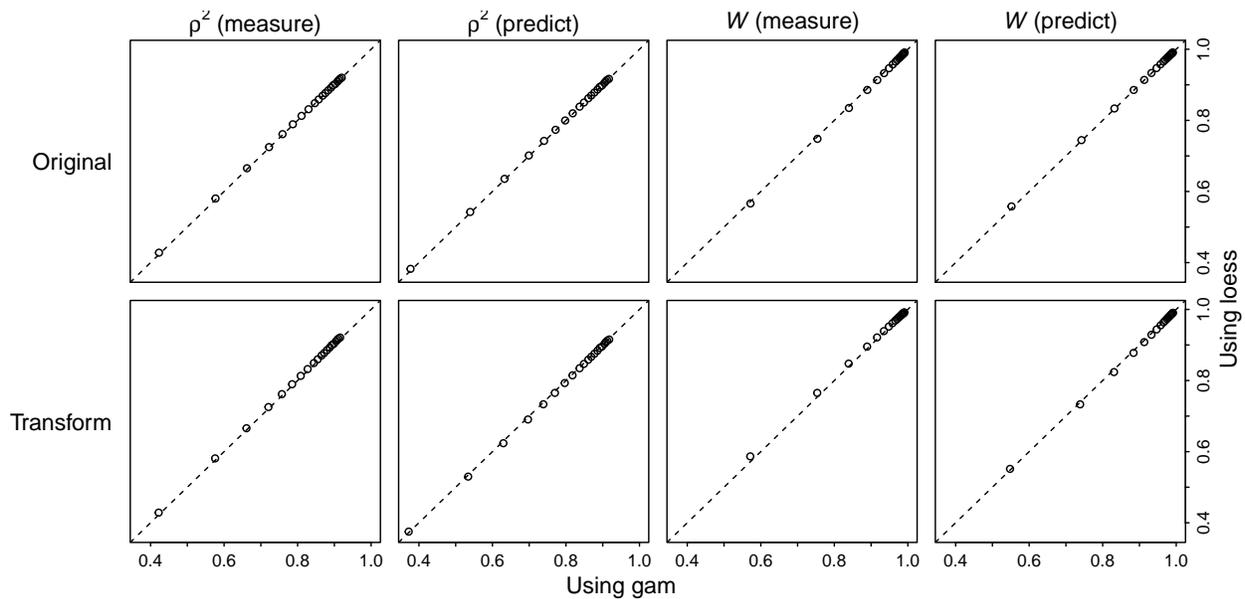

Figure S1: Relationship between different nonparametric regression estimators on reliability estimates. In each panel, regression estimates based on local polynomial regression (obtained using the R function `loess`) are plotted against those based on thin-plate spline regression. The diagonal lines indicating equivalence are displayed as the dashed lines in each panel. Original = when latent scores are original LVs (corresponding to Figure 2B), Transform = when latent scores are percentile ranks of LVs (corresponding to Figure 2C), measure = observed score as outcome, predict = latent score as outcome, $\varrho^2$ = coefficient of determination, and $W$ = coefficient $W$.


 
\end{document}